\documentclass{article}
\usepackage{amsmath}
\usepackage{graphicx}

\title{De Haas-van Alphen effect in two- and quasi two-dimensional 
metals and superconductors}
\author{T. Champel and V.P. Mineev\\
Commissariat \`{a} l'Energie Atomique,  DRFMC/SPSMS\\
17 rue des Martyrs, 38054 Grenoble, France}
\date{August 17, 2000}

\begin{document}

\maketitle

\begin{abstract}
An analytical form of the quantum magnetization oscillations (de 
Haas-van Alphen effect) is derived for two- and quasi two-dimensional 
metals in normal and superconducting mixed states. The theory is 
developed under condition $\mu/\omega_{c}\gg1$ ($\mu$ is the chemical 
potential and $\omega_{c}$ the cyclotron frequency), which is proved 
to be valid for using grand canonical ensemble in the systems of low 
dimensionality. Effects of impurity, temperature, spin-splitting and 
vortex lattice - in the case of superconductors of type II -, are 
taken into account. Contrary to the three dimensional case, the 
oscillations in sufficiently pure systems of low dimensionality and 
at sufficiently low temperatures are characterized by a saw-tooth 
wave form, which smoothened with temperature and concentration of 
impurities growth. In the normal quasi two-dimensional systems, the 
expression for the magnetization oscillations includes an extra 
factor expressed through the transfer integral between the layers. 
The additional damping effect due to the vortex lattice is found. The 
criterion of proximity to the upper critical field for the 
observation of de Haas-van Alphen effect in the superconducting mixed 
state is established.
\end{abstract}

\section{Introduction}
De Haas-van Alphen (dHvA) effect is well known to be a powerful tool 
to get information on the shape of the Fermi surface of normal 
metals. In contrast to three-dimensional (3D) metals where the 
experimental observations have well established  base given by the 
Lifshitz-Kosevich (1956) theory, there is no unanimously 
acknowledged and used theory applicable to 2D or quasi 2D layered 
materials.
Schoenberg (1984) introduced an expression for magnetic 
oscillations in 2D metals where he put phenomenologically the effect 
of temperature. Effect of impurities is not considered, although it 
can be important. There are two essential features of dHvA effect in 
2D case commonly believed differing it from its 3D counterpart :

(i) The sharp saw-tooth like shape of dHvA signal at sufficiently low 
temperatures.

(ii) Strong dependence of the chemical potential on the magnetic 
field, that is  why the proper derivation has to be done in the 
canonical ensemble with fixed number of particles.

To make clear this subject we note that the property (i) takes place 
 in 2D case both at constant number of particles or at constant 
 chemical potential
so long 
we are in the low temperature $T \ll \omega_{c}$ and high purity region 
$\Gamma < \omega_{c}$. Here, $\omega_{c}¥$ is the distance between 
Landau levels, $\Gamma$ is the width of level. As for property (ii) its validity is directly 
related to the concentration of the charge carriers we deal with. For 
low enough concentrations when $\omega_{c} \sim \varepsilon_{F}¥=\frac{\pi 
n}{m^{\ast}}$ ($n$ is the electron density and $m^{\ast}$ is the 
effective mass), the oscillations of the chemical potential $\mu$ with 
magnetic field are important and one has to proceed with the 
calculations in canonical ensemble (see the 
papers (Jauregui {\it et al.} 1990, Harrison{\it et al.} 1996, Grigoriev
and Vagner 1999, Nakano 1998)). On the 
contrary under the condition $\mu \approx \varepsilon_{F}¥=
\frac{\pi n}{m^{\ast}} \gg 
\omega_{c}$ the oscillations of the chemical potential are small and 
one may work making use the grand canonical ensemble as in 3D case.

The analytical theory of dHvA effect in 2D and layered quasi 2D 
metals taking into account the effects of temperature, impurities and 
spin-splitting is developed in the first part of this paper. It is 
proved that the grand canonical ensemble is well accurate for systems 
of low dimensionality in the limit $\mu/\omega_{c} \gg 1$. In pure 
case and at zero temperature the oscillations have the saw-tooth 
shape. The finite temperature as well as the finite amount of 
impurities (even at zero temperature) lead to the natural smoothening 
of the oscillations. The results of calculations (but not the 
calculations themselves) of magnetization oscillations in frame of 
grand canonical ensemble for quasi 2D spinless and perfect crystal 
model can be found in the paper (Nakano 1998). If do not take in mind 
the opposite sign of magnetization (it is probably due to a misprint) 
they coincide with ours taken in the absence of impurities and spin 
paramagnetism. The recent experimental studies (Wosnitsa {\it et al.} 
2000) in several 
organic compounds demonstrate magnetization oscillations behavior 
similar to that should be in the systems with fixed chemical 
potential.

In a parallel direction to the study of dHvA effect in the normal low 
dimensional systems, the possibility of doing quantum oscillations 
measurements in the superconducting mixed state has been recognized 
after its first manifestation by Graebner and 
Robbins (Graebner and Robbins 1976). During the last ten years, dHvA effect has 
already been observed both in mixed state and normal state of many 
types of superconductors like
$NbSe_{2}$ (Haworth {\it et al.} 1999), 
$Nb_{3}Sn$ (Harrison {\it et al.} 1994),
$V_{3}Si$ (Corcoran {\it et al.} 1994), 
$YNi_{2}B_{2}C$ (Goll {\it et al.} 1996, Terashima {\it et al.} 1997),
$CeRu_{2}$ (Hedo {\it et al.} 1998), 
$UPd_{2}Al_{3}$ (Haga {\it et al.} 1999),
$URu_{2}Si_{2}$ (Ohkuni {\it et al.} 1999), 
$\kappa-(BEDT-TTF)_{2}Cu(NCS)_{2}$ (Sasaki {\it et al} 1998).
 Apart the 
low temperature 
$$T< \omega_{c}¥\sim \frac{ eH_{c2}¥}{ m c}
\sim \frac{T_{c}¥^{2}¥}
{\varepsilon_{F}¥ }$$ 
the condition of  observation of dHvA effect in the superconducting 
mixed state consists of ultra purity of the samples 
$$l>\frac{v_{F}¥}{\omega_{c}¥}\sim 
\frac{\varepsilon_{F}¥} { T_{c}¥}
\xi_{0}¥.$$ 
Here
$l$ is the mean free path of quasiparticles,
$\xi_{0}¥$ is the coherence length. Both of these demands 
mean that all enumerated above samples of materials are
ultra clean and at the same time
very strong type-II superconductors with high upper critical field 
value. The latter is provided by the higher critical temperature and 
the lower 
Fermi energy  than
in ordinary type-II superconductors. 

The main observation consists of 
that the frequency of oscillations in the mixed state remains the 
same but the amplitude decreases with decreasing field more rapidly 
than in the normal state. The effect becomes unobservable at fields
several times smaller than $H_{c2}¥$ but still much larger than the 
thermodynamic critical field $H_{c}¥=H_{c2}/\kappa$. Here $\kappa$ 
is the Ginzburg-Landau parameter, which is of the order of 20-30 
typically for observations of dHvA in mixed state materials. That 
means the distance between the vortices is of the order of its core 
diameter in the whole region of the observation. Hence the latter can 
be considered as vicinity of the upper critical field where the 
Abrikosov solution for the vortex lattice serves as a good 
approximation.

Another important 
observation (Goll {\it al.} 1996; 
T.Terashima {\it et al.} 1997)
is that dHvA effect in the 
superconducting mixed state persists below the 
upper critical field not for all but
for the electron orbits with relatively smaller 
radius (or cross section area).

Numerous theoretical studies were done to understand the effect of 
vortex lattice. The essential point here is a nondiagonality 
of the order parameter matrix in the Landau representation preventing 
a simple derivation of the quasiparticle energy spectrum in the mixed 
state. 
The intention to avoid of this problem leads to the idea
to work
with oversimplifyed BCS type spectrum (Miller and Gyorffy  
1995; Dukan and Tesanovich 
1995); Gvozdikov and 
Gvozdikova 1998),
which is valid either in the 
quasiclassical
region  far below of $H_{c2}¥$ where the dHvA effect does not take place 
or in ultra quantum limit with only few filled Landau levels.
In the latter case there are 
however doubts as in the applicability of the BCS theory as in the 
existance of the superconductivity. 

It should be mentioned also an attempt to develope a theory 
for anisotropic superconductors with nodes in the 
gap on the Fermi surface far below $H_{c2}¥$ 
where a vortex core radius is 
much smaller than the distance between vortices , and one can assume 
the space constancy of the order parameter modulus (Gorkov and 
Schrieffer 1998).  As we pointed 
above it does not corresponds to the observations region.

The proper derivation
has been proposed by K.Maki  (1991) and M.Stephen
(1992).
There were established an additional attenuation of the dHvA amplitude in the 
superconducting mixed state. 
The order parameter has 
been determined self-consistently by M.Stephen that has 
allowed to point out 
 the field interval near $H_{c2}¥$
where the effect is accessible for observation. 
The mentioned results were obtained in neglect by the nondiagonal 
elements of the self-energy matrix , or 
sort of random vortex lattice potential approximation has been used.

Somethat later the numerical calculations for two-dimensional electron
system has been performed (Norman and 
MacDonald 1996;
Bruun, Nicos Nicopoulos and Johnson 1997).

Another type of analitical results ( with less rapid 
attenuation of dHvA signal with decreasing of magnetic field in the 
superconducting region) were represented in 
the paper (Zhuravlev et al, 1997), where 
so called strict phase coherence approach has been applied.
Although the latter has definitive 
theoretical interest it seems to be less realistic than the random vortex lattice
potential approximation has been used in the Stephen's paper.

The  treatments  (Stephen 1992, Norman and McDonald 1996, Bruun, 
Nicos Nicopoulos and Johnson 1997, Zhuravlev et al, 1997) were limited by the 
condition $T> \omega_{c}¥/2\pi^{2}¥$ which is necessary for the 
convergency at low temperatures. 
The spreading of the theory on the 
low temperature region $~\omega_{c}¥>\Gamma>T~$ by an introduction of the 
width of the Landau levels $\Gamma$ originating of impurity scattering
has been done 
in the papers  (Vavilov and Mineev 1997; 1998; Mineev 1999).
It was shown that a superconducting state is gapless in the 
mixed state region below the upper critical field if
\begin{equation}
\frac{H_{c2}¥-H}{H_{c2}¥}<\sqrt{\frac{\omega_{c}¥}{\varepsilon_{F}¥}}
\ln \frac{\varepsilon_{F}¥}{\omega_{c}¥},
\label{cond}
\end{equation}
where $\varepsilon_{F}¥/\omega_{c}¥$ is the number of Landau levels 
below the Fermi surface at $H=H_{c2}¥$.
This field interval is negligible for 
any typical type-II superconductor. However for those particular 
ultraclean materials with very high $H_{c2}$ and very small 
$\varepsilon_{F}$, where dHvA effect in the superconducting mixed 
state have been observed, the value of 
$\sqrt {\varepsilon_{F}¥/\omega_{c2}}$ is of the order 
of or even smaller than ten and the  presented theory has sizable 
region of applicability below the upper critical field.
In this region, the oscillatory contribution to the density of states 
and the damping of the amplitude of the magnetization oscillation in 
the superconducting state were found. 

It also important to note that amount of Landau levels below the 
Fermi energy is different for the different bands ($\varepsilon_{F}¥$
in (\ref{cond}) 
should be diminished on the value of energy at the center of the band under 
consideration). That is why the
condition (\ref{cond}) is less restrictive for the electron orbits 
with smaller radius  or cross section area. This fact is in 
correspondence with the observations mentioned above.

In the unconventional 
superconducting states where the lines of extremal cross-section of 
Fermi-surface by the plane perpendicular to magnetic field coincide 
with lines of zeros of the order parameter the amplitude of the 
magnetization oscillations is practically the same as in a normal state. 
In the superconducting state with an other distribution of zeros the 
damping of dHvA oscillations corresponds qualitatively to the ordinary 
superconductivity. 

The further development of the theory of the dHvA effect in the 
superconducting mixed state ,
has been done in the paper  
(Mineev 2000). In three-dimentional isotropic model there 
was analitically derived the Landau 
expansion for the free energy of the superconducting mixed state   
near the upper critical field in powers of the square modulus of
the order parameter averaged over Abrikosov lattice 
with Landau quantization of the quasiparticle energy
taken into consideration.

The condition of the valitity of such the 
expantion were established and  it is given by the 
formula (\ref {cond}). 

As in the normal metal the oscillating with magnetic field terms
represent the tiny corrections to the nonoscillating  free energy
of the superconducting state. However they are fast oscillating 
functions of the ratio $2\pi \varepsilon_{F}¥/\omega_{c}¥$. Becouse of 
that after the differentiation over the magnetic field  $M=- \left( 
\frac{\partial F}{\partial H} \right)_{\mu, T}$  the
oscillating part of the magnetization in the superconducting mixed 
state proves to be even larger than corresponding nonoscillating
part of magnetization.

The analytical theory of the dHvA effect in the superconducting mixed 
state in the vicinity of the upper critical field for 2D and quasi 2D 
layered superconductors is developed in the second part of the 
present article. Our treatment follows the paper (Mineev 2000) where 
the corresponding calculations have been performed for 3D 
superconductors.

\section{De Haas-van Alphen effect in two-dimensional normal metals}

In order to derive the magnetization M, it is necessary to calculate 
first the free energy F which is related to the thermodynamic 
potential $\Omega$ in grand canonical ensemble by\footnote{Planck's 
constant $\hbar$, the Boltzmann's constant $k_{B}$ and the ratio of 
electron charge $|e|$ to the velocity of light c are chosen to be 
unity throughout the paper} 
\begin{equation}
F=\Omega+\mu N \end{equation} with \begin{equation} 
\Omega=-T \int_{0}^{\infty}g(\varepsilon)\ln(1+e^ 
\frac{\mu-\varepsilon}{T})\,d\varepsilon 
\label{potentiel}
\end{equation}
where $\mu$ is the chemical potential of the system, N is the 
number of electrons per unit volume and $g(\varepsilon)$ is the 
density of states in presence of the magnetic field H and the 
impurities. 
\subsection{Density of States}

The latter can be achieved from the imaginary part of the space 
averaged single-electron Green's function $\bar{G}(\varepsilon)$ 
using the relation  
\begin{equation}
g(\varepsilon)=-\frac{1}{\pi} sign(\varepsilon-\mu) \,\Im 
\,\bar{G}(\varepsilon). 
\label{densite} 
\end{equation}
We then follow the proceeding of Bychkov (1961), who has 
dealt with the effect of impurities  in 3D metals under magnetic field using a 
perturbation method. The broadening of each Landau level into the energy 
distribution with Lorenz density of states 
has been found in this paper supposed to be not so bad aproximation in 
2D and quasi 2D cases apparently valid in not very strong magnetic 
fields (see Prange 1987). 
So, we shall use the Green's function 
in the following form 

\begin{equation} 
\bar{G}(\varepsilon)=\sum_{m}\frac{1}{\varepsilon-E_{m}+i\Gamma 
sign(\varepsilon-\mu)} 
\label{Greenmoyen}
\end{equation}
where m specifies the quantum numbers, $\Gamma$ the width of Landau's 
levels due to impurity scattering ($\Gamma=1/2 \tau$ with $\tau$ a 
mean free time of quasiparticles) and  $E_{m}$ the energy spectrum of 
the stationary states in absence of impurities. The quantum numbers 
depend on the dimensionality of the considered system : for the 2D 
case, the set of quantum numbers m includes the magnetic quantum 
number n of the Landau's levels, a wave number $k_{y}$ which is 
connected with the center of the orbit of the electrons, and the spin 
$\sigma =\pm 1$ ; for the 3D case, we have in addition the momentum  
$k_{z}$. The field $\mathbf{H}$ is directed along the z axis, that is 
in 2D case perpendicular to the conducting plane. The energies are : 

\begin{equation} 
E_{n, k_{z}, 
\sigma}^{3d}=\omega_{c}(n+\frac{1}{2})+\frac{k_{z}^{2}}{2m}+\sigma 
\mu_{e}H ,
\end{equation}
\begin{equation}
 E_{n, \sigma}^{2d}=\omega_{c}(n+\frac{1}{2})+\sigma \mu_{e}H 
\end{equation}
One needs to remember that all the energies should be 
counted from the energy $\varepsilon_{b}$ at the center of the considered 
band. Keeping 
in mind this circumstance we shall shift the chemical potential in 
final formulas on this value.
The expression for the single Green's function for a given spin 
$\sigma$ becomes in the 2D case 

\begin{equation} 
\bar{G}^{\sigma}(\varepsilon)=
N_{0}\omega_{c}\sum_{n=0}^{\infty}\frac{1}{\varepsilon-\omega_{c}(n+\frac{1}{2})-\sigma 
\mu_{e}H+i\Gamma sign(\varepsilon-\mu)}.
\label{Green} 
\end{equation}
Here $N_{0}=\frac{m}{2 \pi}$ is the normal metal electron density of 
states per one spin projection and $\mu_{e}$ is the electron's 
magnetic moment. The expression (\ref{Green}) is 
divergent. In fact, it is the case only for the real part of 
$\bar{G}^{\sigma}(\varepsilon)$ ; the imaginary part, which we have 
to calculate next, is convergent 
\begin{equation}
\Im\,\bar{G}^{\sigma}(\varepsilon)=N_{0}\omega_{c} 
\sum_{n=0}^{+\infty} \frac{-\Gamma 
sign(\varepsilon-\mu)}{(\varepsilon-\omega_{c}(n+\frac{1}{2})-\sigma 
\mu_{e}H)^{2}+ \Gamma^{2}}.
\end{equation}
The Poisson's summation formula 
$$\sum_{n=0}^{\infty}f(n)=\sum_{l=-\infty}^{+\infty} 
\int_{a}^{+\infty} f(t) e^{-2\pi i lt}\,dt$$
where $a$ is a number 
between -1 and 0, can now be used properly. Writing 
$\varepsilon=\xi+\mu$, changing variable $t$ to 
$x=t-\frac{\mu}{\omega_{c}}$, and then setting the lower limit of the 
integral equal to $-\infty$ since we consider the limit 
$\frac{\mu}{\omega_{c}}\gg 1$, we get 

\begin{equation}
\Im\,\bar{G}^{\sigma}(\varepsilon)=- \pi N_{0} sign\xi 
+\Im\,\bar{G}_{osc}^{\sigma}(\varepsilon) 
\label{Im}
\end{equation}
with
\begin{equation}
\Im\,\bar{G}_{osc}^{\sigma}(\varepsilon)=-N_{0}\omega_{c} \Gamma 
sign\xi \sum_{l=-\infty, \neq 0}^{+\infty} e^{-2 \pi i l 
\frac{\mu}{\omega_{c}}} \int_{-\infty}^{+\infty} \frac{e^{-2 \pi i l 
x} \,dx}{(\xi-\omega_{c}(x+\frac{1}{2})-\sigma 
\mu_{e}H)^{2}+\Gamma^{2}}.
\label{G}
\end{equation}
The first term in the right hand side of equation (\ref{Im}) comes 
from the contribution of $l=0$ in the sum and corresponds to the 
result in absence of field. The other contributions $l \neq 0$ give 
an oscillatory part depending on the field. The remaining integration 
is quite straightforward by a residue calculation and yields 

\begin{equation}
\Im\,\bar{G}_{osc}^{\sigma}(\varepsilon)=2 \pi sign(\varepsilon-\mu) 
N_{0} \sum_{l=1}^{+\infty} (-1)^{l+1} \cos\left({2 \pi l 
\frac{(\varepsilon - \sigma \mu_{e} H)}{\omega_{c}}}\right) \, e^{-2 
\pi l \frac{\Gamma}{\omega_{c}}}.
\end{equation}
From relation (\ref{densite}), we get directly the searched density 
of states, after summing up the two spin states  

\begin{equation} 
g^{2d}(\varepsilon)=2N_{0} \left[ 1+2 \sum_{l=1}^{+\infty} (-1)^{l} 
\cos(2 \pi l \frac{\varepsilon}{\omega_{c}}) \cos(2 \pi l 
\frac{\mu_{e}H}{\omega_{c}}) \, e^{-2 \pi l 
\frac{\Gamma}{\omega_{c}}} \right].
\end{equation}

\subsection{Chemical potential}

Making use of equation (\ref{potentiel}), the oscillatory part of the 
thermodynamic potential $\Omega$ is performed by an integration by 
parts and by the change of variable $x=\frac{(\varepsilon-\mu)}{T}$ 

\begin{equation}
\Omega^{2d}_{osc}=2N_{0} T \omega_{c} \sum_{l=1}^{+\infty} 
\frac{(-1)^{l+1}}{l} \, e^{-2 \pi l \frac{\Gamma}{\omega_{c}}} \cos(2 
\pi l \frac{\mu_{e}H}{\omega_{c}}) \frac{1}{\pi} \int_{- 
\frac{\mu}{T}}^{+\infty} \frac{\sin(2\pi l 
\frac{(\mu+Tx)}{\omega_{c}})}{1+e^{x}} \,dx.
\end{equation}
The lower limit of the integral may be set equal to $-\infty$ since 
in our investigation we have the condition $\mu \gg  T$. Using the 
value of the integral  
$$ \int_{-\infty}^{+\infty} \frac{e^{i \alpha 
y}}{1+e^{y}} \,dy =- \frac{i \pi}{\sinh(\alpha \pi)}$$ 
we find at last 

\begin{equation}
\Omega^{2d}_{osc}=2N_{0} T \omega_{c} \sum_{l=1}^{+\infty} 
\frac{(-1)^{l}}{l} \, e^{-2 \pi l \frac{\Gamma}{\omega_{c}}} \cos(2 
\pi l \frac{\mu}{\omega_{c}}) \frac{\cos(2 \pi l 
\frac{\mu_{e}H}{\omega_{c}})}{\sinh(\frac{2 \pi^{2} lT}{\omega_{c}})}.
\end{equation}
The nonoscillatory part of the thermodynamic potential gives on its 
side 
\begin{equation}
\Omega^{2d}_{0}=-N_{0} \mu^{2}. 
\end{equation}
The number of electrons is defined by $N=- \left( \frac{\partial 
\Omega}{\partial \mu} \right)_{T, H}$ and thus  

\begin{equation}
N=2 N_{0} \mu +4\pi N_{0} T \sum_{l=1}^{+\infty} (-1)^{l} 
\frac{\cos(2 \pi l \frac{\mu_{e}H}{\omega_{c}})}{\sinh(\frac{2 
\pi^{2} lT}{\omega_{c}})} \sin(2 \pi l \frac{\mu}{\omega_{c}}) \, 
e^{-2 \pi l \frac{\Gamma}{\omega_{c}}} 
\label{N}.
\end{equation}
This formula determines of the number of particles as the function
of $(H,T)$ at fixed $\mu$. On the other hand one can consider it as 
the equation for the chemical potential as the function of $(H,T)$
at fixed number of particles.
Recognizing  the Fermi energy 
\begin{equation}
\varepsilon_{F}=\frac{N}{2N_{0}}
\end{equation}
one can rewright this equation as

\begin{equation}
\mu=\varepsilon_{F}¥+\frac{\omega_{c}¥}{\pi}
\sum_{l=1}^{+\infty} \frac{(-1)^{l+1}}{l} \frac{\lambda_{l}}{\sinh 
\lambda_{l}} \sin(2 \pi l \frac{\mu}{\omega_{c}}) \cos(2 \pi l 
\frac{\mu_{e}H}{\omega_{c}}) \, e^{-2 \pi l 
\frac{\Gamma}{\omega_{c}}}
\label{mu}
\end{equation}
with 
$$\lambda_{l}=\frac{2 \pi^{2} lT}{\omega_{c}}.$$
We see that at $\varepsilon_{F}¥\gg \omega_{c}¥$ the second oscillating 
term represent the small correction to the const value of chemical 
potential. On the other hand the expression (\ref {G}) has been 
obtained in neglect of terms of the order of $\omega_{c}¥/\mu$. That 
is why it would be out of limits our accuracy to keep the difference
between the chemical potential and Fermi energy in the right hand side
of (\ref {mu}). 
So, finally we get
\begin{equation}
\mu=\varepsilon_{F}¥+\frac{\omega_{c}¥}{\pi}
\sum_{l=1}^{+\infty} \frac{(-1)^{l+1}}{l} \frac{\lambda_{l}}{\sinh 
\lambda_{l}} \sin(2 \pi l \frac{\varepsilon_{F}¥}{\omega_{c}}) \cos(2 \pi l 
\frac{\mu_{e}H}{\omega_{c}}) \, e^{-2 \pi l 
\frac{\Gamma}{\omega_{c}}}.
\label{mu1}
\end{equation}
It is worth to remind here that in real crystal one should to shift 
all the energies on the energy in center of band: $\mu\to 
\mu-\varepsilon_{b}¥,~ \varepsilon_{F}¥\to \varepsilon_{F}¥-\varepsilon_{b}¥$.

At zero temperature, ignoring for brevity the spin splitting
and summing up over $l$, we find a simple expression
\begin{equation}
\mu=\varepsilon_{F}+\frac{\omega_{c}¥}{\pi} 
\arctan \left( \frac{ \sin(2 \pi \frac{\varepsilon_{F}¥}{\omega_{c}})}{\cos(2 \pi 
\frac{\varepsilon_{F}¥}{\omega_{c}})+e^{2 \pi \frac{\Gamma}{\omega_{c}}}} \right)
\end{equation}
which is obviously has at $\Gamma=0$ saw-tooth shape.
At any finite $\Gamma$, the 
oscillations of the chemical potential are smooth. For instance for 
$2 \pi \Gamma/\omega_{c}\geq 1$ 
\begin{equation}
\mu=\varepsilon_{F}+\frac{\omega_{c}¥}{\pi} 
\sin(2 \pi \frac{\varepsilon_{F}¥}{\omega_{c}})
e^{-2 \pi \frac{\Gamma}{\omega_{c}}}.
\end{equation}


On this stage one should stress ones more that all the results has 
been obtained here at the condition 
$\mu/\omega_{c} \gg 1$. 

\subsection{Magnetization}

The magnetization $M=- \left( 
\frac{\partial \Omega}{\partial H} \right)_{\mu, T}$ 
can be found from  the relation (14) . Keeping only 
the preponderant terms and ignoring as before the small difference 
between $\mu$ and $\varepsilon_{F}¥$ in the arguments of oscillating 
harmonics 
\footnote {The restoration of the oscillating difference $\mu-\varepsilon_{F}¥$
would be resulted in the additional frequences in Fourier spectrum of 
the magnetization which certainly present when the ratio 
$\omega/\varepsilon_{F}¥$ is not small. 
One should stress however that both formulas for the 
chemical potential and for the magnetization are derived here by making use
the expression for the density of states have been found in the neglect of terms 
of the order of $\omega_{c}¥/\varepsilon_{F}¥$ and  keeping of such 
the terms  in the oscillating arguments would be out the limits of accuracy we used.}, 
we obtain

\begin{equation}
M^{2d}_{n,osc}=\sum_{l=1}^{+\infty}M^{2d}_{l}=N_{0}\frac{2}{\pi} \frac{\omega_{c}}{H}
\varepsilon_{F}¥ 
\sum_{l=1}^{+\infty} \frac{(-1)^{l+1}}{l} \frac{\lambda_{l}}{\sinh 
\lambda_{l}} \sin(2 \pi l \frac{\varepsilon_{F}¥}{\omega_{c}}) \cos(2 \pi l 
\frac{\mu_{e}H}{\omega_{c}}) \, e^{-2 \pi l 
\frac{\Gamma}{\omega_{c}}}.  
\label{aimantation} 
\end{equation}

This formula without impurities factor corresponds to the intuitive 
formula given by Schoenberg (1984). 
After (\ref{mu1}) and (\ref{aimantation}) we notice 
that  at $\varepsilon_{F}¥\gg \omega_{c}¥$ the magnetization is directly related to the 
oscillating part of $\mu$  by
\begin{equation}
M^{2d}_{n,osc}=\frac{N}{H}\mu_{osc}=
M_{0}\frac{\mu_{osc}}{\omega_{c}¥}.
\end{equation}
This result, which is valid at any temperature smaller than the 
$\varepsilon_{F}¥$, generalizes  that was given in the recent 
paper (Itskovsky, Maniv and Vagner 2000) for the range of temperature $T \leq 
\omega_{c}$.

For T=0, this expression  is simplified since a summation over the 
integers $l$ is possible. Without spin-splitting, we have  

\begin{equation}
M^{2d}_{n,osc}=\frac{2 N_{0}}{\pi} \frac{\omega_{c}}{H} \varepsilon_{F}¥ 
\arctan \left( 
\frac{ \sin(2 \pi \frac{\varepsilon_{F}¥}{\omega_{c}})}{\cos(2 \pi 
\frac{\varepsilon_{F}¥}{\omega_{c}})+e^{2 \pi \frac{\Gamma}{\omega_{c}}}} \right) 
\label{aim}
\end{equation}
and with spin-splitting 
\begin{equation}
M^{2d}_{n,osc}=\frac{N_{0}}{\pi} \frac{\omega_{c}}{H} \varepsilon_{F}¥ \arctan \left( 
\frac{ \sin(4 \pi \frac{\varepsilon_{F}¥}{\omega_{c}})+2 \sin(2 \pi 
\frac{\varepsilon_{F}¥}{\omega_{c}}) \cos(2 \pi \frac{\mu_{e} H}{\omega_{c}}) e^{2 
\pi \frac{\Gamma}{\omega_{c}}}}{\cos(4 \pi \frac{\varepsilon_{F}¥}{\omega_{c}})+ 2 
\cos(2 \pi \frac{\varepsilon_{F}¥}{\omega_{c}}) \cos(2 \pi \frac{\mu_{e} 
H}{\omega_{c}}) e^{2 \pi \frac{\Gamma}{\omega_{c}}} +e^{4 \pi 
\frac{\Gamma}{\omega_{c}}}} \right).
\end{equation}
For $2 \pi\Gamma/\omega_{c} \geq 1$, this yields 
\begin{equation}
M^{2d}_{n,osc}=\frac{2 N_{0}¥}{\pi}\frac{\omega_{c}}{H}\varepsilon_{F}¥ 
\sin(2\pi\frac{\varepsilon_{F}¥}{\omega_{c}})\cos(2\pi\frac{\mu_{e}H}{\omega_{c}})\, 
e^{-2\pi\frac{\Gamma}{\omega_{c}}}.
\end{equation}

\section{De Haas-van Alphen effect in qua\-si two-di\-men\-sio\-nal 
nor\-mal metals}

In layered metals, the simplest spectrum for electrons is
\begin{equation}
 E_{n, k_{z}, \sigma}=\omega_{c}(n+\frac{1}{2})-2t\cos(k_{z}d)+\sigma 
\mu_{e}H. 
\label{spectre}
\end{equation}
The parameter t is the transfer integral along the z axis, d is the 
distance between the layers. Here the field $\mathbf{H}$ is 
perpendicular to the layers. More generally, in a tilted magnetic 
field, the spectrum energy is quantized as (Yamaji 1989) 
\begin{equation}
 E_{n, k_{z}, \sigma}=\omega_{c} \cos\theta \,(n+\frac{1}{2})-2t 
J_{0}(k_{F}d\tan\theta) \cos(k_{z}d)+\sigma \mu_{e}H 
\end{equation}
where $J_{0}$ is the Bessel's function of zero order and $k_{F}$ is 
the momentum at the Fermi's surface. For convenience, we will work 
with spectrum (\ref{spectre}). The corresponding generalization for 
the case of finite tilting is easy with simple substitution of the 
angle dependent parameters. 
According to (\ref{Greenmoyen}), the space averaged Green's function 
becomes henceforth 

\begin{equation} 
\bar{G}^{\sigma}(\varepsilon)=\frac{N_{0} \omega_{c}}{2 \pi d} 
\sum_{n=0}^{\infty} \int_{-\pi}^{\pi} 
\frac{\,d\psi}{\varepsilon-\omega_{c}(n+\frac{1}{2})-\sigma 
\mu_{e}H+2t\cos\psi+i\Gamma sign(\varepsilon-\mu)}.
\end{equation}
Here we consider that $t \leq$ or $\sim \omega_{c}$ so that roughly 
speaking only the modulation of one Landau level is taken into 
account. For $t \gg \omega_{c}$, many Landau levels play a role and 
we can do the following approximation 
$$-(n+\frac{1}{2})+\frac{2t}{\omega_{c}} \cos\psi \approx 
-(n+\frac{1}{2})+\frac{2t}{\omega_{c}} (1-\frac{(k_{z}d)^{2}}{2}).$$ 
In this limit, we are again in the 3D case.
Next, we proceed exactly like previously using the Poisson's 
summation formula. The density of states includes then an additional 
Bessel function factor 

\begin{equation} 
g^{q2d}(\varepsilon)=\frac{2 N_{0}}{d} \left[ 1+2 
\sum_{l=1}^{+\infty} (-1)^{l} \cos(2 \pi l 
\frac{\varepsilon}{\omega_{c}}) J_{0}(2 \pi l \frac{2t}{\omega_{c}})  
\cos(2 \pi l \frac{\mu_{e}H}{\omega_{c}})\,e^{-2 \pi l 
\frac{\Gamma}{\omega_{c}}} \right]
\end{equation}
with  $$J_{0}(z)=\frac{1}{2\pi} \int_{-\pi}^{+\pi} e^{2\pi i z \cos 
\psi} \,d\psi$$
and the oscillatory part of the thermodynamic potential becomes 

\begin{equation}
\Omega^{q2d}_{osc}=2\frac{N_{0}}{d} T \omega_{c} \sum_{l=1}^{+\infty} 
\frac{(-1)^{l}}{l} \frac{\cos(2 \pi l 
\frac{\mu}{\omega_{c}})}{\sinh(\lambda_{l})} J_{0}(2 \pi l 
\frac{2t}{\omega_{c}})  \cos(2 \pi l \frac{\mu_{e}H}{\omega_{c}}) \, 
e^{-2 \pi l \frac{\Gamma}{\omega_{c}}}.
\end{equation}
Thus 

\begin{eqnarray}
M^{q2d}_{n,osc}=\frac{N_{0}}{d} \frac{2}{\pi} \frac{\omega_{c}}{H} \mu 
\sum_{l=1}^{+\infty} \frac{(-1)^{l+1}}{l}  \frac{\lambda_{l}}{\sinh 
\lambda_{l}} \cos(2 \pi l \frac{\mu_{e}H}{\omega_{c}})\,e^{-2 \pi l 
\frac{\Gamma}{\omega_{c}}} \nonumber \\ \times \left\{\sin(2 \pi l 
\frac{\mu}{\omega_{c}})  J_{0}(2 \pi l \frac{2t}{\omega_{c}})+ 
\frac{2t}{\mu} \cos(2 \pi l \frac{\mu}{\omega_{c}})  J_{1}(2 \pi l 
\frac{2t}{\omega_{c}})\right\}
\end{eqnarray}
where $J_{1}(z)=-J^{\prime}_{0}(z)$ is the Bessel's function of first 
order.

Assuming that the transfer energy $t$ is much smaller than the Fermi 
energy $\varepsilon_{F}$, we obtain 

\begin{equation}
M^{q2d}_{n,osc}=\frac{N_{0}}{d} \frac{2}{\pi} \frac{\omega_{c}}{H} \mu 
\sum_{l=1}^{+\infty} \frac{(-1)^{l+1}}{l} \frac{\lambda_{l}}{\sinh 
\lambda_{l}} \sin(2 \pi l \frac{\mu}{\omega_{c}})  J_{0}(2 \pi l 
\frac{2t}{\omega_{c}}) \cos(2 \pi l 
\frac{\mu_{e}H}{\omega_{c}})\,e^{-2 \pi l \frac{\Gamma}{\omega_{c}}}. 
\label{aimantationq}
\end{equation}

Here we remind that in reality the chemical potential $\mu$ has to be 
reduced on the energy of the band center $\mu\to  \mu-\varepsilon_{b}
\approx\varepsilon_{F}¥-\varepsilon_{b}$. Without spin-splitting 
and impurities factor, this result is the same as given 
in (Nakano 1998) except the opposite sign probably due to a 
misprint.
The extra factor $J_{0}(2 \pi l \frac{2t}{\omega_{c}})$ in this 
expression induces a rich behavior for the manifestly field dependent 
amplitude of the harmonics. In formal limit $2t \gg \omega_{c}$, 
$$J_{0}(z) \sim \sqrt{\frac{2}{\pi z}} \cos(z-\frac{\pi}{4})$$ and 
thus we return back to expression closed to Lifshitz-Kosevich 
result (Lifshits and Kosevich 1956) for the amplitude of the harmonics. For $2t \ll 
\omega_{c}$, $J_{0}(z) \sim 1$ and we find again the 2D expression 
(\ref{aimantation}).

\section{De Haas-van Alphen effect in mixed state of two- and quasi 
two-dimensional superconductors}

In this section, we want to determine the effect of the 
low-dimensionality on the additional damping factor due to the vortex 
lattice. In this purpose, we follow the work (Mineev 2000) (and also 
take the same notations) where the theory of dHvA effect for 3D 
metals in the superconducting mixed state has been developed, and 
adapt it to 2D and quasi 2D systems. It should be mentioned that 
under 2D systems we shall imply 3D layered crystals with negligibly 
small interaction between layers, such that the mixed state in the 
magnetic field perpendicular to the layers represents Abrikosov but 
not Pearl vortex lattice. The treatment of the problem is carried out 
in the framework of the Gorkov's formalism for a conventional 
superconductor.
\subsection{Free energy}
The free energy density in mixed state near the upper critical field 
$H_{c2}$ is developed in powers of the square modulus of the order 
parameter averaged over the Abrikosov's lattice (Mineev 2000)
\begin{equation}
F_{s}-F_{n}=\alpha \Delta^{2}+\frac{\beta}{2} \Delta^{4}
\end{equation}
where
\begin{equation}
\alpha=\frac{1}{g} -\int e^{-\frac{H \rho^{2}}{2}} K_{2}(\mathbf{R})  
\,d \mathbf{R} \label{alpha}
\end{equation}
and
\begin{equation}
\beta=\frac{1}{V} \int f^{\ast}(\mathbf{r_{1}})  f(\mathbf{r_{2}})  
f^{\ast}(\mathbf{r_{3}})  f(\mathbf{r_{4}}) K_{4}(\mathbf{r_{1}}, 
\mathbf{r_{2}}, \mathbf{r_{3}}, \mathbf{r_{4}}) \,d \mathbf{r_{1}} 
\,d \mathbf{r_{2}} \,d \mathbf{r_{3}} \,d \mathbf{r_{4}}.
\end{equation}
Here g is the constant of the pairing interaction, 
$\rho^{2}=R^{2}-Z^{2}$, 
$$f(\mathbf{r})=2^{1/4} 
\sum_{\nu=integer}\exp\left(i\frac{2\pi\nu}{a}y-\left(\frac{x}{\lambda}+
\frac{\pi\nu}{a}\lambda 
\right)^{2} \right)$$ 
is the Abrikosov square 
lattice solution for the order parameter where the elementary cell edge $a$ is such that 
$a^{2}=\pi \lambda^{2}$ and $\lambda=H^{-1/2}$ is the magnetic length 
and $\Delta$ is the order parameter amplitude.

The functions $K_{2}$ and $K_{4}$ have the following expressions : 
\begin{equation}
K_{2}(\mathbf{R})=\frac{1}{2} \sum_{\sigma= \pm 1} T \sum_{\nu} 
\tilde{G}^{\sigma}(\mathbf{R}, \tilde{\omega}_{\nu}) 
\tilde{G}^{-\sigma}(\mathbf{R}, -\tilde{\omega}_{\nu}),
\end{equation}
\begin{equation}
G^{\sigma}(\mathbf{r}_{1}, \mathbf{r}_{2}, \tilde{\omega}_{\nu})=\exp 
\left( i \int_{\mathbf{r}_{1}}^{\mathbf{r}_{2}} 
\mathbf{A}(\mathbf{s})\,d\mathbf{s} \right) 
\tilde{G}^{\sigma}(\mathbf{r}_{1}-\mathbf{r}_{2}, 
\tilde{\omega}_{\nu}),
\end{equation}
\begin{eqnarray}
K_{4}(\mathbf{r_{1}}, \mathbf{r_{2}},\mathbf{r_{3}}, 
\mathbf{r_{4}})=\frac{1}{2} \sum_{\sigma= \pm 1} T \sum_{\nu} 
G^{\sigma}(\mathbf{r_{1}}, \mathbf{r_{2}}, \tilde{\omega_{\nu}})  
G^{-\sigma}(\mathbf{r_{2}}, \mathbf{r_{3}}, -\tilde{\omega_{\nu}})  
G^{\sigma}(\mathbf{r_{3}}, \mathbf{r_{4}}, \tilde{\omega_{\nu}}) 
\nonumber \\ \times G^{-\sigma}(\mathbf{r_{1}}, \mathbf{r_{4}}, 
-\tilde{\omega_{\nu}}),
\end{eqnarray}
\begin{equation}
\tilde{\omega}_{\nu}=\omega_{\nu}+\Gamma sign\omega_{\nu}, 
\hspace{2em} \omega_{\nu}=\pi T(2\nu+1).
\end{equation}
To obtain this formula, it is assumed that the magnetic field is 
uniform and coincides with external field, which is a good 
approximation in the vicinity of $H_{c2}$ for superconductors with 
large Ginzburg-Landau parameter.

The coefficients $\alpha$ and $\beta$, which depend on the 
dimensionality of the system by way of the Green's functions, are 
calculated respectively in appendix A and in appendix B in the 
semi-classical approximation $\frac{\mu}{\omega_{c}} \gg 1$. They 
consist of a smooth function and a fast oscillating function of the 
magnetic field :
\begin{equation}
\alpha(H, T)=\bar{\alpha}(H, T) + \alpha_{osc}(H, T),
\end{equation}
\begin{equation}
\beta(H, T)=\bar{\beta}(H, T) + \beta_{osc}(H, T).
\end{equation}
For a two-dimensional system, we find 
\begin{equation}
\bar{\alpha}^{2d}(H, T)=\frac{N_{0}}{2} 
\frac{H-\bar{H}_{c2}^{2d}(T)}{H_{c2o}^{2d}}
\end{equation} where $\bar{H}_{c2}(T)$ is the upper critical field at 
low temperatures averaged over the oscillations,
\begin{eqnarray}
\alpha_{osc}^{2d}(H, T)=-4 \pi^{3/2} N_{0} \frac{T}{\sqrt{\mu 
\omega_{c}}} \sum_{l=1}^{+\infty} (-1)^{l} \cos(2 \pi l 
\frac{\mu}{\omega_{c}}) \cos( 2 \pi(l+2) \frac{\mu_{e}H}{\omega_{c}}) 
\nonumber \\ \times \frac{e^{-2 \pi (l+2) 
\frac{\Gamma}{\omega_{c}}}}{\sinh \lambda_{l+2}} 
\end{eqnarray}
and 
 \begin{equation}
\bar{\beta}^{2d}(H,T=0)=N_{0} \frac{\omega_{c}}{16 \pi \mu} 
\frac{\Gamma^{2}-(\mu_{e}H)^{2}}{(\Gamma^{2}+(\mu_{e}H)^{2})^{2}} 
\approx N_{0} \frac{\omega_{c}}{16 \pi \mu \Gamma^{2}} 
\end{equation} which is valid for $\mu_{e}H<\Gamma<\omega_{c}$,
\begin{equation}
\beta_{osc}^{2d}(H, T=0)=N_{0} \frac{\omega_{c}}{2 \pi \mu 
\Gamma^{2}}  \sum_{l=1}^{+\infty} (-1)^{l} \cos(2 \pi l 
\frac{\mu}{\omega_{c}}) I(2 \pi l \frac{\Gamma}{\omega_{c}}) e^{-2\pi 
l \frac{\Gamma}{\omega_{c}}}
\end{equation}
where $I(x)=\int_{0}^{\infty} \left[ 
\frac{1}{(y+1)^{3}}+\frac{x}{(1+y)^{2}} \right] e^{-xy} \,dy$ is a 
bounded function.
Minimization of the free energy over $\Delta$ yields
\begin{equation}
F_{s}=F_{n}-\frac{\alpha^{2}}{2\beta}.
\end{equation}
\subsection{Magnetization}
Magnetization is given by $M_{s}=-\frac{\partial F_{s}}{\partial H}$. 
In the differentiation we keep only the fast oscillating terms 
\begin{equation}
M_{s, osc} \simeq M_{n, osc}+ \frac{\bar{\alpha}}{\bar{\beta}} 
\frac{\partial \alpha_{osc}}{\partial H}-\frac{\bar{\alpha}^{2}}{2 
\bar{\beta}^{2}} \frac{\partial \beta_{osc}}{\partial H}.
\end{equation}
Here $M_{n, osc}$ is the oscillating part of the normal metal 
magnetization calculated previously (see equation 
(\ref{aimantation})). Like in the 3D case (Mineev 2000), at $4 \pi 
\Gamma \sim \omega_{c}$ the following inequality takes place
\begin{equation}
|M_{n, osc}| > |\frac{\bar{\alpha}}{\bar{\beta}} \frac{\partial 
\alpha_{osc}}{\partial H}| > |-\frac{\bar{\alpha}^{2}}{2 
\bar{\beta}^{2}} \frac{\partial \beta_{osc}}{\partial H}|
\end{equation}
until the new more restrictive than in the 3D case (see (\ref{cond})) condition :
\begin{equation}
\frac{\bar{H}_{c2}(T)-H}{H_{c2o}}<\left(\frac{\omega_{c}}{\mu}\right)^{1/2}. 
\label{condition}
\end{equation}

With the previous condition we keep also only the first two terms and 
obtain for the oscillating part of the magnetization 
\begin{equation}
M_{s, osc}^{2d}=M_{n, osc}^{2d}+\frac{\bar{\alpha}}{\bar{\beta}} 
\frac{\partial \alpha_{osc}}{\partial H}=\sum_{l=1}^{+\infty} 
M_{nl}^{2d} M_{sl}^{2d},
\label{1}
\end{equation}
where 
\begin{equation}
M_{nl}^{2d}=N_{0}\frac{2}{\pi} \frac{\omega_{c}}{H} \mu  
\frac{(-1)^{l+1}}{l} \frac{\lambda_{l}}{\sinh 
\lambda_{l}} \sin(2 \pi l \frac{\mu}{\omega_{c}}) \cos(2 \pi l 
\frac{\mu_{e}H}{\omega_{c}}) \, e^{-2 \pi l 
\frac{\Gamma}{\omega_{c}}}  
\label{tation} 
\end{equation}
and
\begin{eqnarray}
M_{sl}^{2d}=1-\sqrt{\pi}\left(\frac{4 \pi 
\Gamma}{\omega_{c}}\right)^{2}\sqrt{\frac{\mu}{\omega_{c}}} 
\frac{\bar{H}_{c2}-H}{H_{c2o}} \frac{l^{2}\lambda_{l+2} 
\sinh\lambda_{l}}{(l+2)\lambda_{l} \sinh\lambda_{l+2}} \nonumber \\ 
\times \frac{\cos(2\pi(l+2)\frac{\mu_{e}H}{\omega_{c}})}{\cos(2\pi 
l\frac{\mu_{e}H}{\omega_{c}})} e^{-\frac{4 \pi\Gamma}{\omega_{c}}} 
\label{aimantationsupra}
\end{eqnarray}

with 
$$\lambda_{l}=\frac{2 \pi^{2} lT}{\omega_{c}}.$$

All the previous calculations can be done in the same way for a 
quasi-two dimensional system. We give only the final result  

\begin{equation}
M_{sl}^{q2d}=M_{sl}^{2d}.
\end{equation}

The field interval (\ref{condition}) in which the results (\ref{1}),
 (\ref{tation}), (\ref{aimantationsupra}) are valid is tiny for 
an ordinary type-II superconductor. However for those particular 
ultraclean materials with very high $H_{c2}$ and very small 
$\varepsilon_{F}$, where dHvA effect in the superconducting mixed 
state have been observed (see introduction), the value of 
$\sqrt {\mu/\omega_{c2}}$  is of the order of ten and
the theory presented here has noticable 
region of applicability below the upper critical field. 

As the final remarque, we have to remember that $\mu$ is the chemical 
potential less the energy band $\varepsilon_{b}$ so that the limit 
(\ref{condition}) is different for different bands and less 
restrictive for the bands with smallest value of 
$\mu-\varepsilon_{b}\approx \varepsilon_{F}¥-\varepsilon_{b}$.

\section{Conclusion}
The grand canonical ensemble theory of de Haas-van Alphen effect in 
the normal and superconducting states of 2D and quasi 2D metals valid 
at $\mu \gg \omega_{c}$ is developed taking into consideration 
finite temperature, level broadening due to impurity scattering, spin 
paramagnetic splitting and additional suppression due to 
inhomogeneous order parameter distribution in the mixed state. The 
main results are represented by the expressions (\ref{aimantation}), 
(\ref{aimantationq}) and (\ref{1}), (\ref{tation}), (\ref{aimantationsupra}) 
for the 
magnetization amplitude oscillations.

The straightforward 
generalization to the multiband case is obtained by the summation of 
these expressions over several energetic bands. 

The limits of 
observation of dHvA effect near the upper critical field 
(\ref{condition}) in the superconducting mixed state will be of 
course different for different energetic bands being less restrictive 
for the bands with smallest orbit area $\sim (\varepsilon_{F}¥ -\varepsilon_{b})$.

\section*{Acknowledgments}
We express our gratitudes to Yu.A.Bychkov, I.D.Vagner, P.D.Grigoriev 
and E. Steep for discussions and help in navigation in the de Haas
van Alphen sea.

\appendix
\section*{Appendix A : Calculation of $\alpha$}
In this Appendix, we perform the calculation of the coefficient 
$\alpha(H,T)$ in the quasi-classical limit. Effects of quantization 
on the upper critical field have already been studied by Gruenberg 
and Gunther (1968) for a 3D system. These authors give a 
method for the calculation of the corrections in the linearized gap 
equation, which can be adapted to our 2D case. According to equation 
(\ref{alpha}) we have
\begin{equation}
\alpha(H,T)=\frac{1}{g}-T \sum_{\nu=0}^{+\infty} 
\bar{S}_{\tilde{\omega}_{\nu}}
\end{equation}
where 
\begin{equation}
\bar{S}_{\tilde{\omega}_{\nu}}=2 \Re \int e^{-\frac{H \rho^{2}}{2}} 
\tilde{G}^{\sigma}(\mathbf{R},\tilde{\omega}_{\nu} ) 
\tilde{G}^{-\sigma}(\mathbf{R},-\tilde{\omega}_{\nu}) \,d\mathbf{R}. 
\label{Somega}
\end{equation} 
In the 2D case, the Green's function $\tilde{G}^{\sigma}(\mathbf{R}, 
\tilde{\omega}_{\nu})$ in the representation $\mathbf{R}$, after the 
integration over orbit centers (Dworin 1966), is expressed as 

\begin{equation}
\tilde{G}^{\sigma}(\mathbf{R}, \tilde{\omega}_{\nu})=N_{0} \omega_{c} 
e^{-t/2} \sum_{r=0}^{+\infty} \frac{L_{r}(t)}{i 
\tilde{\omega}_{\nu}-\xi_{r}+\sigma\mu_{e}H}
\end{equation}
where $L_{r}(t)$ is the Laguerre's polynomial of order $r$, $t=H 
\rho^{2}/2$, and $$\xi_{r}=(r+\frac{1}{2}) \omega_{c}-\mu.$$
Using the relation between Laguerre's polynomials 
$$\int_{0}^{+\infty} e^{-2t} 
L_{n}(t)L_{m}(t)\,dt=\left(\frac{1}{2}\right)^{n+m+1} 
\frac{(m+n)!}{m!\,n!}$$ 
we obtain
\begin{equation}
\bar{S}_{\tilde{\omega}_{\nu}}=N_{0} \omega_{c} 
\Re\sum_{r,l=0}^{+\infty}  \frac{(r+l)!}{r!\,l!} 
\frac{(1/2)^{r+l}}{(i\tilde{\omega}_{\nu}+
\xi_{r}-\mu_{e}H)(-i\tilde{\omega}_{\nu}+\xi_{l}+\mu_{e}H)}.
\end{equation}
Following Gruenberg and Gunther (1968), we do the Gaussian 
approximation $$\left(\frac{1}{2}\right)^{r+l} \frac{(r+l)!}{r!\,l!} 
\approx \frac{e^{-(r-l)^{2}/4r}}{\left( \pi r \right)^{1/2}}$$ valid 
for $\mu/\omega_{c} \gg 1$. Then we make use of the Poisson's formula 
which gives us three kinds of terms :
\begin{equation}
\bar{S}_{\tilde{\omega}_{\nu}}=\sum_{n,m=-\infty}^{+\infty} 
S_{\tilde{\omega}_{\nu}}^{nm} \label{S},
\end{equation}
\begin{equation}
S_{\tilde{\omega}_{\nu}}^{nm} \approx (-1)^{n+m} N_{0} \omega_{c} \Re 
\int_{0}^{\infty}\!\!\int_{0}^{\infty}dy \frac{1}{\sqrt{\pi x}} 
\frac{e^{2 \pi i(nx-my)-(x-y)^{2}/4x}\,dx 
\,dy}{(i\tilde{\omega}_{\nu}+\xi_{x}-\mu_{e}H)(-i\tilde{\omega}_{\nu}+\xi_{y}+\mu_{e}H)} 
\end{equation}
with $$\xi_{x}=x\omega_{c}-\mu. $$

In the first term obtained for $n=m=0$, we transform from integration 
over $x$ and $y$ to integration over the coordinates of the 
two-dimensional vectors $\mathbf{q}$ and $\mathbf{q'}$, so that 
$\omega_{c}x=q^{2}/2m$, $\omega_{c}y=q^{\prime 2}/2m$. Since q and q' 
are very close (and of order of $k_{F}$), we have $$(x-y)^{2}/4x 
\approx (q-q')^{2}/2H$$
After using the following approximative relation 
$$\frac{e^{-(q-q')^{2}/2H}}{\sqrt{qq'/2H}} \approx 
\frac{1}{\sqrt{\pi}} \int_{0}^{2 \pi} 
e^{-(\mathbf{q}-\mathbf{q'})^{2}/2H}\,d\theta_{\mathbf{q}\mathbf{q'}}$$ 
changing variable $\mathbf{q'}$ to 
$\mathbf{Q}=\mathbf{q}-\mathbf{q'}$ and neglecting the term 
$Q^{2}/2m$ in the energy $\xi_{q'}$, we obtain 
\begin{equation}
S_{\tilde{\omega}_{\nu}}^{00} \approx \frac{N_{0} \omega_{c}}{2 
\pi^{2} H^{2}} \Re \int \! \! d\mathbf{Q} \,e^{-Q^{2}/2H} \int\! 
\frac{d 
\mathbf{q}}{(i\tilde{\omega}_{\nu}+\xi_{q}-\mu_{e}H)(-i\tilde{\omega}_{\nu} 
+ \xi_{q}+\mu_{e}H-\mathbf{Q} \cdot \frac{\mathbf{q}}{m})}.
\end{equation}
Since $q$ is very close to $k_{F}$, the term $q/m$ is approximate by 
the Fermi velocity $v_{F}$. The integration over $\mathbf{q}$ is 
changed to an integral over energy and angle $\theta$. The first 
integration over energy gives 
\begin{equation}
S_{\tilde{\omega}_{\nu}}^{00}=\frac{N_{0}}{\pi H} \Re\int d 
\mathbf{Q} \, e^{-Q^{2}/2H} \int_{0}^{2\pi} \frac{d \theta 
}{2\tilde{\omega}_{\nu}+2 i \mu_{e}H-iv_{F}Q\cos\theta}.
\end{equation}
The second integration over the angle is performed by introducing the 
complex variable $z=e^{i\theta}$. The corresponding contour is the 
unity circle. We find 
\begin{equation}
S_{\tilde{\omega}_{\nu}}^{00}=\frac{2N_{0}}{H} \Re \int  
\frac{e^{-Q^{2}/2H}\,d 
\mathbf{Q}}{\sqrt{(v_{F}Q)^{2}+4(\tilde{\omega}_{\nu}+i\mu_{e}H)^{2}}}.
\end{equation} 
Rewriting this expression as 
\begin{equation}
S_{\tilde{\omega}_{\nu}}^{00}=2 \pi N_{0} \zeta \Re 
\int_{0}^{+\infty}  \frac{e^{-\zeta x}\,dx}{\sqrt{\kappa+x}} 
\label{fin}
\end{equation} 
where $\zeta=1/\varepsilon_{F}\omega_{c}$  ($\varepsilon_{F}$ is the 
Fermi level) and $\kappa=(\tilde{\omega}_{\nu}+i\mu_{e}H)^{2}$, then 
we calculate 
$$S^{00}=\frac{T}{N_{0}} \sum_{\nu=0}^{+\infty} S^{00}_{\tilde{\omega}_{\nu}}.$$ 
Using the following relation valid at 
low temperatures 
\begin{equation}
2 \pi T \sum_{\nu} F(\omega_{\nu})=\int_{0}^{+\infty}F(\omega) 
\,d\omega +\frac{\pi^{2}}{6}T^{2} \left(\frac{\partial F}{\partial 
\omega} \right)_{\omega=0} \label{temperature}
\end{equation}
we get
\begin{equation}
\frac{1}{gN_{0}}-S^{00}=\frac{\zeta}{2} \int_{0}^{+\infty} \ln x \, 
e^{-\zeta x}\,dx+\frac{\pi^{2}}{6}T^{2} \zeta \sqrt{\kappa_{0}} 
\int_{0}^{+\infty} \frac{e^{-\zeta x} 
\,dx}{(x+\kappa_{0})^{3/2}}-\ln\frac{\pi T_{c}}{\gamma}.
\end{equation}
Here $\kappa_{0}=\Gamma+i\mu_{e}H$. In the first term in the 
right-hand side of equation (\ref{temperature}), we neglect $\Gamma$ 
and $\mu_{e}H$. To avoid the divergence, a cut-off $\omega_{0}$ is 
introduced ; this latter vanishes with subtraction of the term 
$\frac{1}{N_{0}g}=\ln(\frac{2\gamma \omega_{0}}{\pi T_{c}})$ where 
$C=\ln\gamma$ is the Euler constant. Taking notes of the value of the 
integrals (Gradshteyn and Ryzhik 1980) :
$$\zeta\int_{0}^{+\infty} \ln x \, e^{-\zeta x}\,dx=-C-\ln\zeta$$ and 
$$\int_{0}^{+\infty} \frac{e^{-\zeta x} 
\,dx}{(x+\kappa_{0})^{3/2}}=\frac{2}{\sqrt{\kappa_{0}}}-2\sqrt{\pi\zeta} 
e^{\kappa_{0}\zeta}\left(1-\frac{2}{\sqrt{\pi}}\int_{0}^{\sqrt{\zeta\kappa}}e^{-t^{2}}\,dt\right) 
\approx \frac{2}{\sqrt{\kappa_{0}}}$$ we obtain 
\begin{equation}
\frac{1}{g}-T \sum_{\nu=0}^{+\infty} 
S^{00}_{\tilde{\omega}_{\nu}}=\frac{N_{0}}{2} 
\left\{\ln(\frac{H}{H_{c2o}})+\frac{2\pi^{2}}{3}\frac{T^{2}}{\mu\omega_{c}} 
\right\}
\end{equation}
with $H_{c2o}=\frac{2 \pi^{2} T_{c}^{2}}{\gamma v_{F}^{2}}$.
The temperature dependent term is neglected further in so far as we 
work at very low-temperatures.

For the second term obtained for $n=m\neq 0$ in (\ref{S}), we follow 
the steps leading to equation (\ref{fin}). We approximate the 
additional term in the exponent by $2i \pi n(x-y) \approx 2 i \pi n 
v_{F}\cos\theta$. Thus we find for $n>0$, 
$S_{\tilde{\omega}_{\nu}}^{nn}=0$ and for $n<0$ 
\begin{equation}
S_{\tilde{\omega}_{\nu}}^{nn}=2 \pi N_{0} \zeta \Re \, 
e^{-4\pi|n|\frac{(\tilde{\omega}_{\nu}+i\mu_{e}H)}{\omega_{c}}} 
\int_{0}^{+\infty}  \frac{e^{-\zeta x}\,dx}{\sqrt{\kappa+x}}.
\end{equation} 
Using the value of the integral (Gradshteyn and Ryzhik 1980) $$\int_{0}^{+\infty} 
\frac{e^{-\zeta x}\,dx}{\sqrt{\kappa+x}}=\sqrt{\frac{\pi}{\zeta}}\, 
e^{\zeta 
\kappa}\left(1-\frac{2}{\sqrt{\pi}}\int_{0}^{\sqrt{\zeta\kappa}}e^{-t^{2}}\,dt\right) 
\approx \sqrt{\frac{\pi}{\zeta}}$$ for $\zeta\kappa \ll 1$, we get  
\begin{equation}
S_{\tilde{\omega}_{\nu}}^{nn}=\frac{2 \pi^{3/2} 
N_{0}}{\sqrt{\mu\omega_{c}}}\Re \, e^{-4 \pi n 
\frac{(\tilde{\omega}_{\nu}+i \mu_{e}H)}{\omega_{c}}}.
\end{equation}

The third and last term in (\ref{S}) containing the summation over $n 
\neq m$ is evaluated only for nonzero values of $n$ and $m$ because 
$S_{\tilde{\omega}_{\nu}}^{n0}$ and  $S_{\tilde{\omega}_{\nu}}^{0n}$ 
are negligibly small. We approximate $(\pi 
x)^{-1/2}\exp[-(x-y)^{2}/4x]$ by $(\omega_{c}/\pi \mu)^{1/2}$ in 
order to have two uncoupled integrations. These are nonzero only for 
$n$ and $m$ negative and yield for a pair $(n,m)$  
\begin{equation}
S_{\tilde{\omega}_{\nu}}^{nm}+S_{\tilde{\omega}_{\nu}}^{mn}=(-1)^{n+m}\frac{8 
\pi^{3/2} N_{0}}{\sqrt{\mu\omega_{c}}}\Re \, e^{-2 \pi |n+m| 
(\frac{\tilde{\omega}_{\nu}+i \mu_{e}H}{\omega_{c}})} \cos(2 \pi 
(n-m)\frac{\mu}{\omega_{c}}).
\end{equation}
Reassembling all these different contributions, we have
\begin{equation}
\alpha(H, T)=\frac{N_{0}}{2} \left\{ 
\frac{H-H_{c2o}}{H_{c2o}}+\left(\frac{T}{T_{c}}\right)^{2} 
S_{0}-2\sqrt{\pi}\sqrt{\frac{\omega_{c}}{\mu}} 
S_{1}-8\sqrt{\pi}\sqrt{\frac{\omega_{c}}{\mu}} S_{2} \right\}
\end{equation}
where 
\begin{equation}
S_{0}=\frac{2 \pi^{2}T_{c}^{2}}{3 \mu \omega_{c}},
\end{equation}
\begin{equation}
S_{1}=\frac{2\pi T}{\omega_{c}} \Re \sum_{n=1}^{+\infty} 
\sum_{\nu=0}^{+\infty} e^{-4 \pi n 
(\frac{\tilde{\omega}_{\nu}+i\mu_{e}H}{\omega_{c}})},
\end{equation}
\begin{equation}
S_{2}=\frac{2\pi T}{\omega_{c}} \Re \sum_{n=1}^{+\infty} 
\sum_{m>n}^{+\infty} \sum_{\nu=0}^{+\infty} (-1)^{n+m} \cos(2 \pi 
(n-m)\frac{\mu}{\omega_{c}}) e^{-2 \pi (n+m) 
(\frac{\tilde{\omega}_{\nu}+i\mu_{e}H}{\omega_{c}})} \label{S2}.
\end{equation}
The term $S_{1}$ is the same as in the 3D case (Mineev 2000). We 
also do the same transformation and include it in the upper critical 
field at low temperatures averaged over the oscillations.
Like in (Mineev 2000), we simplify the term $S_{2}$ in (\ref{S2}) 
changing summation variables to $n$ and $m-n=l$, and keeping 
reasonably only the first term in the sum over $n$ 
\begin{equation}
S_{2}=\frac{\pi T}{\omega_{c}} \sum_{l=1}^{+\infty} 
(-1)^{l}\frac{e^{-2 \pi (l+2) \frac{\Gamma}{\omega_{c}}}}{\sinh 
\lambda_{l+2}} \cos(2\pi (l+2) \frac{\mu_{e}H}{\omega_{c}}) \cos(2 
\pi l \frac{\mu}{\omega_{c}}).
\end{equation}
This is the term which is taken into account in $\alpha_{osc}$.

In the quasi 2D case, we can follow exactly the same previous 
proceeding to calculate the coefficient $\alpha$. The Green's 
function is expressed in a mixt representation $(x,y,k_{z})$ 
\begin{equation}
\tilde{G}^{\sigma}(x,y,k_{z}, \tilde{\omega}_{\nu})=N_{0} \omega_{c} 
e^{-t/2} \sum_{r=0}^{+\infty} \frac{L_{r}(t)}{i 
\tilde{\omega}_{\nu}-(r+\frac{1}{2})\omega_{c}+2t\cos(k_{z}d)+\mu+\sigma\mu_{e}H}.
\end{equation}
In equation (\ref{Somega}), in addition to the integration over the 
two-dimensional vector $\mathbf{R}$ there is an additional  $k_{z}$ 
integration.
At last, we obtain 
\begin{equation}
\alpha^{q2d}(H, T)=\frac{N_{0}}{2d} \left\{ 
\frac{H-H_{c2o}}{H_{c2o}}+\left(\frac{T}{T_{c}}\right)^{2} 
S_{0}-2\sqrt{\pi}\sqrt{\frac{\omega_{c}}{\mu}} 
S_{1}-8\sqrt{\pi}\sqrt{\frac{\omega_{c}}{\mu}} S_{2} \right\}
\end{equation}
where
$$S_{0}^{q2d}=\frac{S_{0}^{2d}}{d},$$
$$S_{1}^{q2d}=\frac{S_{1}^{2d}}{d},$$
$$S_{2}^{q2d}=\frac{\pi T}{\omega_{c}} \sum_{l=1}^{+\infty} 
(-1)^{l}\frac{e^{-2 \pi (l+2) \frac{\Gamma}{\omega_{c}}}}{\sinh 
\lambda_{l+2}} \cos(2\pi (l+2) \frac{\mu_{e}H}{\omega_{c}}) J_{0}(2 
\pi l \frac{2t}{\omega_{c}}) \cos(2 \pi l \frac{\mu}{\omega_{c}}).$$
Therefore, the oscillating part of $\alpha$ includes an additional 
Bessel function factor analogous to that in quasi 2D normal metal 
expressions.

\section*{Appendix B : Calculation of $\beta$}
The calculation of the coefficient $\beta$ is performed in the 
magnetic sublattices' representation (Vavilov and Mineev 1997) 
\begin{eqnarray}
\beta(H,T)=\frac{T}{2} \sum_{\sigma \pm 1}\sum_{v} \sum_{n, n', m, 
m'} G^{-\sigma}(\xi_{n}, -\tilde{\omega}_{\nu}) G^{\sigma}(\xi_{m}, 
\tilde{\omega}_{\nu}) G^{-\sigma}(\xi_{n'}, -\tilde{\omega}_{\nu}) 
\nonumber \\ \times G^{\sigma}(\xi_{m'}, \tilde{\omega}_{\nu}) 
F_{mm'}^{nn'} \label{beta}
\end{eqnarray}
where  $F_{mm'}^{nn'}=\int \frac{d\mathbf{q}}{(2 \pi)^{2}}   
f_{nm}(\mathbf{q}) f^{\ast}_{nm'}(\mathbf{q})  f_{n'm}(\mathbf{q}) 
f^{\ast}_{n'm'}(\mathbf{q})  $ are the matrix elements of the 
Abrikosov's function f. Keeping only the main contribution to the 
expression (\ref{beta}) given by the diagonal terms $n=n'=m=m'$, and 
since $ F_{nn}^{nn} \approx H/(2 \pi)^{2}n$ (Vavilov and Mineev 1997), we have
\begin{equation}
\beta(H, T)=\frac{H}{2 (2 \pi)^{2}} \sum_{\sigma \pm 1} T \sum_{\nu} 
\sum_{n=0}^{+\infty} \frac{[G^{\sigma}(\xi_{n}, \tilde{\omega}_{\nu}) 
G^{-\sigma}(\xi_{n}, -\tilde{\omega}_{\nu})]^{2}}{n}.
\end{equation}
The use of the Poisson's summation formula to transform the summation 
over $n$ leads to a smooth non-oscillating field dependent part 
and to a fast oscillating field dependent part of$\beta$. The smooth part is 
obtained by the substitution of the summation by the integration with 
$$\frac{H}{2\pi} \sum_{n=0}^{+\infty}=\frac{H}{2\pi} 
\int_{n=0}^{+\infty}=N_{0} \int \,d\xi$$ and $n \approx 
\mu/\omega_{c}$ :
\begin{equation}
\bar{\beta}(H,T)=\frac{N_{0} \omega_{c}}{4 \pi \mu} \sum_{\sigma \pm 
1} T \sum_{v} \int_{-\infty}^{+\infty} \frac{\,d\xi}{[(i 
\tilde{\omega}_{\nu}-\xi+\sigma \mu_{e}H)(-i 
\tilde{\omega}_{\nu}-\xi-\sigma \mu_{e}H)]^{2}}.
\end{equation}
The integration over energy $\xi$ gives 
\begin{equation}
\bar{\beta}(H, T)=N_{0} \frac{\omega_{c}}{4 \mu} T 
\sum_{\nu=0}^{+\infty} \frac{\tilde{\omega}^{3}_{\nu} -3 
\tilde{\omega}_{\nu} ( 
\mu_{e}H)^{2}}{(\tilde{\omega}^{2}_{\nu}+(\mu_{e}H)^{2})^{2}}.
\end{equation}
This expression is then performed at $T=0\, K$ to allow the change of 
the summation to the integration. We find
\begin{equation}
\bar{\beta}(H,0)=N_{0} \frac{\omega_{c}}{16 \pi \mu} 
\frac{\Gamma^{2}-(\mu_{e}H)^{2}}{(\Gamma^{2}+(\mu_{e}H)^{2})^{2}}.
\end{equation}

The oscillating part of $\beta(H,T)$ is given by  
\begin{eqnarray}
\beta_{osc}(H,T)=\frac{N_{0} \omega_{c}}{2\pi \mu} \Re \sum_{\sigma 
\pm 1} T \sum_{\nu} \sum_{l=1}^{+\infty} (-1)^{l} \, e^{2 \pi i 
l\frac{\mu}{\omega_{c}}}\nonumber \\ \times 
\int_{-\infty}^{+\infty}\frac{e^{-2 \pi i l \frac{\xi}{\omega_{c}}}\, 
d\xi}{(i \tilde{\omega}_{\nu}-\xi+ \sigma 
\mu_{e}H)^{2}(-i\tilde{\omega}_{\nu}-\xi-\sigma \mu_{e}H)^{2}}.
\end{eqnarray}
The integration over $\xi$ presents no difficulty 
\begin{eqnarray}
\beta_{osc}(H,T)=\frac{N_{0}\omega_{c}}{\mu} \sum_{l=1}^{+\infty} 
(-1)^{l} \cos(2 \pi l\frac{\mu}{\omega_{c}}) \Re \, T 
\sum_{\nu=0}^{+\infty}e^{-2 \pi 
l(\frac{\tilde{\omega}_{\nu}+i\mu_{e}H}{\omega_{c}})} \nonumber \\ 
\times \left\{\frac{1}{(\tilde{\omega}_{\nu}+i\mu_{e}H)^{3}}+\frac{2 
\pi l}{\omega_{c}(\tilde{\omega}_{\nu}+i\mu_{e}H)^{2}} \right\}.
\end{eqnarray}
At $T=0$, the summation over $\nu$ is replaced by the integration, 
which yields at last 
\begin{equation}
\beta_{osc}^{2d}(H, T=0)=\frac{N_{0}\omega_{c}}{2 \pi \mu 
\Gamma^{2}}  \sum_{l=1}^{+\infty} (-1)^{l} \cos(2 \pi l 
\frac{\mu}{\omega_{c}}) I(2 \pi l \frac{\Gamma}{\omega_{c}}) e^{-2\pi 
l \frac{\Gamma}{\omega_{c}}}
\end{equation}
where 
$$I(x)=\int_{0}^{\infty} \left[ 
\frac{1}{(y+1)^{3}}+\frac{x}{(1+y)^{2}} \right] e^{-xy} \,dy.$$

For a quasi 2D system, the same calculations yields
\begin{equation}
\bar{\beta}^{q2d}(H,0)=\frac{\bar{\beta}^{2d}(H,0)}{d}
\end{equation}
and like in the oscillating part of $\alpha$, an additional Bessel's 
function factor appears the oscillating part of $\beta$ 
\begin{equation}
\beta_{osc}^{q2d}(H,0)=\frac{N_{0}}{d} \frac{\omega_{c}}{2 \pi \mu 
\Gamma^{2}}  \sum_{l=1}^{+\infty} (-1)^{l} \cos(2 \pi l 
\frac{\mu}{\omega_{c}}) I(2 \pi l \frac{\Gamma}{\omega_{c}}) J_{0}(2 
\pi l \frac{2t}{\omega_{c}}) \, e^{-2\pi l \frac{\Gamma}{\omega_{c}}}.
\end{equation}

\section*{References}
Bruun G.M., Nicopoulos V. Nicos and Johnson N.F., 1997, {\em 
Phys. Rev. B} \textbf{56}, 809.\\
Bychkov Yu.A., 1961, {\em Sov. Phys. JETP} \textbf{12}, 
977.\\
Corcoran R., Harrison N., Hayden S.M.,  Meeson P., 
Springford M., van der Wel P.J., 1994, {\em Phys. Rev. Lett.} \textbf{72}, 
701.\\
Dukan S. and Tesanovich Z.,1995,{\em Phys. Rev. Lett.} \textbf{74}, 
2311.\\
Dworin L., 1966, {\em Ann. Phys. (N.Y.)} \textbf{38}, 431.\\
Gradshteyn I.S. and J.M. Ryzhik, 1980, {\em Table of Integrals, 
Series and Products}. Academic Press Inc., New York.\\
Graebner J.E., Robbins M., 1976, {\em Phys. Rev. Lett.} 
\textbf{36}, 422.\\
Goll G.,  Heinecke M.,  Jansen A.G.M., Joss W.,  
Nguyen L.,  Steep E.,  Winzer K.,  Wyder P., 1996, {\em Phys. Rev. B} 
\textbf{53}, R8871.\\
Gor'kov L.P. and Schrieffer, 1998,{\em Phys. Rev. Lett.} \textbf{80}, 
3360.\\ 
Grigoriev P.D. and Vagner I.D., 1999, {\em JETP Lett.} 
\textbf{69}, 156.\\
Gruenberg L.W. and L. G\"{u}nther L., 1968, {\em Phys. Rev.} 
\textbf{176}, 606.\\
Gvozdikov V.M. and Gvozdikova M.V., 1998,{\em Phys. Rev. 
B} \textbf{58}, 8716.\\
Haga Y., Y. Inada, K. Sakurai, Y. Tokiwa, E. Yamamoto, 
T. Honma, Y. Onuki, 1999, {\em J. Phys. Soc. Jpn.} \textbf{68}, 342.\\
Harrison N., Hayden S.M., Meeson P., Springford M., 
van der Wel P.J., Menovsky A.A., 1994, {\em Phys. Rev. B} \textbf{50}, 
4208.\\
Harrison N., Bogaerts R., Reinders P.H.P., Singleton J., Blundell 
S.J., Herlach F., 1996, {\em Phys. Rev. B} \textbf{54}, 9977.\\
Haworth C.J., Hayden S.M., Janssen T.J.B.M.,  
Meeson P., Springford M.,  Wasserman A., 1999, {\em Physica B} 
\textbf{259-261}, 1066.\\
Hedo M., Inada Y., Sakurai K., Yamamoto E., Haga Y., 
Onuki Y., Yakahashi S., Higuchi M.,  Maehira T., Hasegawa A., 1998, {\em 
Phil. Mag. B} \textbf{77}, 975.\\
Itskovsky M.A., Maniv T., Vagner I.D., 2000, {\em Phys. 
Rev. B} \textbf{61}, 14616.\\
Jauregui K., Marchenko V.I. and Vagner I.D., 1990, {\em Phys. Rev. B} 
\textbf{41}, 12922.\\
Lifshits I.M. and Kosevich A.M., 1956, {\em Sov. Phys. JETP} 
\textbf{2}, 636.\\
Maki K., 1991, {\em Phys. Rev. B} \textbf{44}, 2861.\\
Miller P. and Gy\"{o}rffy B.L., 1995, {\em J. Phys. : 
Condens. Matter} \textbf{7}, 5579.\\
Mineev V.P., 1999, Physica B \textbf {259}-\textbf {261}, 1072.\\
Mineev V.P., 2000, {\em Phil. Mag. B} \textbf{80}, 307.\\ 
Nakano M., 1998, {\em J. Phys. Soc. Jpn.} \textbf{68}, 
1801.\\
Norman M.R. and Mac Donald A.H., 1996, {\em Phys. Rev. B} 
\textbf{54}, 4239.\\
Ohkuni H., Inada Y., Tokiwa Y., Sakurai K., 
Settai R., Honma T., Haga Y., Yamamoto E., Onuki Y.,  Yamamagi H.,  
Takahashi S.,  Yanagisawa T., 1999, {\em Phil. Mag. B} \textbf{79}, 
1045.\\
Prange R.E., 1987, Chapters 1 and 3 in the book "The Quantum Hall effect",
ed. by R.E.Prange and S.M. Girvin, Springer Verlag, New York, 1987.\\
Sasaki T., Biberacher W., Neumaier K., Hehn W.,  
Andres K., Fukase T., 1998, {\em Phys. Rev. B} \textbf{57}, 10889.\\
Schoenberg D., 1984, {\em J. Low Temp. Phys.} 
\textbf{56}, 417.\\
Stephen M.J., 1992, {\em Phys. Rev. B} \textbf{45}, 5481.\\ 
Terashima T., Haworth C., Takeya H.,  Uji S.,  
Aoki H., Kadowaki K., 1997, {\em Phys. Rev. B} \textbf{56}, 5120.\\
Vavilov M.G. and Mineev V.P., 1997, {\em Sov. Phys. JETP} 
\textbf{85}, 1024.\\
Vavilov M.G. and Mineev V.P., 1998, {\em Sov. Phys. JETP}, \textbf {86}, 
1191.\\ 
Wosnitza J., Wanka S., Hagel J., Balthes E.,  
Harrison N., Schlueter J.A., Kini A.M., Geser U., Mohtasham J.,  
Winter R.W., Gard G.L., 2000, {\em Phys. Rev. B} \textbf{61}, 7383.\\
Yamaji K., 1989, {\em J. Phys. Soc. Jpn.} \textbf{58}, 
1520.\\
Zhuravlev M.G.,  Maniv T.,  Vagner I.D., Wyder P., 1997, 
{\em Phys. Rev. B} \textbf{56}, 14693.

\end{document}